\documentclass[doublecol]{epl2}

\usepackage{epsfig}

\title{Dynamics of Neural Networks with Continuous Attractors}

\author{C. C. Alan Fung\inst{1}
\and K. Y. Michael Wong\inst{1}
\and Si Wu\inst{2}}
\shortauthor{Fung \etal}

\institute{
\inst{1} Department of Physics,
Hong Kong University of Science and Technology, 
Clear Water Bay, Hong Kong, China\\
\inst{2} Department of Informatics, University of Sussex, 
Brighton, United Kingdom
}

\abstract{
We investigate the dynamics of continuous attractor neural networks (CANNs). 
Due to the translational invariance of their neuronal interactions,
CANNs can hold a continuous family of stationary states. 
We systematically explore how their neutral stability 
facilitates the tracking performance of a CANN,
which is believed to have wide applications in brain functions. 
We develop a perturbative approach 
that utilizes the dominant movement of the network stationary states 
in the state space. 
We quantify the distortions of the bump shape during tracking, 
and study their effects on the tracking performance. 
Results are obtained on the maximum speed 
for a moving stimulus to be trackable, 
and the reaction time to catch up an abrupt change in stimulus.}

\begin{document}

\maketitle

Understanding how the dynamics of a neural network 
is shaped by the network structure, 
and consequently facilitates the functions implemented by the neural system, 
is at the core of using physical models 
to elucidate brain functions~\cite{Dayan}. 
Traditional models of attractor neural networks (ANNs), 
such as those based on the Hopfield model~\cite{Hopfield}, 
are powerful for elucidating the computation 
and the memory retrieving processes in the brain. 
When external noisy inputs are presented, 
the dynamics of ANNs will reach an attractor 
that is highly correlated with the memories stored in the network. 
Each attractor has its own basin of attraction 
well separated from the others. 
These models successfully describe the behavior of associative memories, 
and stimulated an upsurge of studies 
on the retrieval behavior of the models~\cite{Domany}.
However, other than memory retrieval, 
many other computational issues of the brain 
have not been analyzed to the same extent.

Recently, a new type of attractor networks, 
called continuous attractor neural networks (CANNs), 
has received considerable attention 
(see, e.g.,~\cite{Amari,Sompolinsky,Zhang,Ermentrout,
Latham,Pinto,Laing,Seung,Wang,Trappenberg,Wu1,Coombes1,
Tsodyks,Coombes2,Wu2,McNaughton,Folias}). 
These networks possess a translational invariance of the neuronal interactions.
As a result, they can hold a family of stationary states 
which can be translated into each other 
without the need to overcome any barriers. 
Thus, in the continuum limit, 
they form a continuous manifold in which the system is neutrally stable, 
and the network state can translate easily 
when the external stimulus changes continuously. 
Beyond pure memory retrieval, 
this endows the neural system with a tracking capability.

The tracking dynamics of a CANN has been investigated
by several authors in the literature 
(see, e.g., \cite{Sompolinsky,Zhang,Seung,Trappenberg,Wu1,Folias}) 
These studies have shown that a CANN has the capacity 
of tracking a moving stimulus continuously
and that this tracking property can well justify some brain functions. 
Despite these successes, however, 
a detailed analysis of the tracking behaviors of a CANN is still lacking.
These include, for instance, 
1) the conditions under which a CANN can successfully track a moving stimulus, 
2) the distortion of the shape of the network state during the tracking, and 
3) the effects of these distortions on the tracking speed. 
In this Letter we will report, as far as we know, the first 
systematic study on these issues.

We will use a simple, analytically-solvable, 
CANN model as the working example.
The form of the network state is reminiscent of those of solitons 
in nonlinear dynamics. 
We display clearly how the dynamics of a CANN 
is decomposed into different distortion modes, 
corresponding to, respectively, 
changes in the height, position, width and skewness 
of the network state. 
We then demonstrate which of them dominates 
the tracking behaviors of the network.
In order to solve the dynamics which is 
otherwise extremely complicated for a large recurrent network, 
we develop a time-dependent perturbation method 
to approximate the tracking performance of the network. 
The solution is expressed in a simple closed-form, 
and we can approximate the network dynamics 
up to an arbitory accuracy depending on the order of perturbation used. 
We expect that our method will provide a useful tool 
for the theoretical studies of CANNs. 
Our work generates new predictions 
on the tracking behaviors of CANNs, namely, 
the maximum tracking speed to moving stimuli, 
and the reaction time to sudden changes in external stimuli, 
both are testable by experiments. 

Specifically, we consider a one-dimensional continuous stimulus 
being encoded by an ensemble of neurons. 
For example, the stimulus may represent 
the moving direction, the orientation, or a general
continuous feature of an external object. 
Let $U(x,t)$ be the synaptic input at time $t$ 
to the neurons with preferred stimulus of real-valued $x$. 
We will consider stimuli and responses with correlation length $a$ 
much less than the range of $x$, 
so that the range can be effectively taken to be $(-\infty,\infty)$.
The firing rate $r(x,t)$ of these neurons 
increases with the synaptic input, 
but saturates in the presence of a global activity-dependent inhibition. 
A solvable model that captures these features is given by
\begin{equation}
	r(x,t)=\frac{U(x,t)^2}{1+k\rho\int dx'U(x',t)^2},
\end{equation}
where $\rho$ is the neural density, 
and $k$ is a small positive constant 
controlling the strength of global inhibition. 
The dynamics of the synaptic input $U(x,t)$ 
is determined by the external input $I_{\rm ext}(x,t)$, 
the network input from other neurons, and its own relaxation. 
It is given by
\begin{equation}
	\tau\frac{dU(x,t)}{dt}
	=I_{\rm ext}(x,t)+\rho\int dx'J(x,x')r(x',t)-U(x,t),
\label{eq:dyn}
\end{equation}
where $\tau$ is the time constant, 
which is typically of the order 1 ms, 
and $J(x,x')$ is the neural interaction from $x'$ to $x$. 
The key characteristic of CANNs is 
the translational invariance of their neural interactions. 
In our solvable model, 
we choose Gaussian interactions with a range $a$, namely,
\begin{equation}
	J(x,x')=\exp[-(x-x')^2/(2a^2)]J/\sqrt{2\pi a^2}.
\label{eq:interaction}
\end{equation}
CANN models with other neural interactions and inhibition mechanisms 
have been studied~\cite{Amari,Sompolinsky,Zhang,Latham,Wang}. 
However, our model has the advantage 
of permitting a systematic perturbative improvement. 
Nevertheless, the final conclusions of our model 
are qualitatively applicable to general cases (to be further discussed at the end of
the paper).

We first consider the intrinsic dynamics of the CANN model 
in the absence of external stimuli. 
For $0<k<k_c\equiv \rho J^2/(8\sqrt{2\pi}a)$, 
the network holds a continuous family of stationary states, which are 
\begin{equation}
	\tilde U(x|z)=U_0\exp\left[-\frac{(x-z)^2}{4a^2}\right],
\end{equation}
where $U_0=[1+(1-k/k_c)^{1/2}]J/(4\sqrt\pi ak)$. 
These stationary states are translationally invariant among themselves 
and have the Gaussian bumped shape peaked at arbitrary positions $z$. 

\begin{figure}
\centerline{\epsfig{file=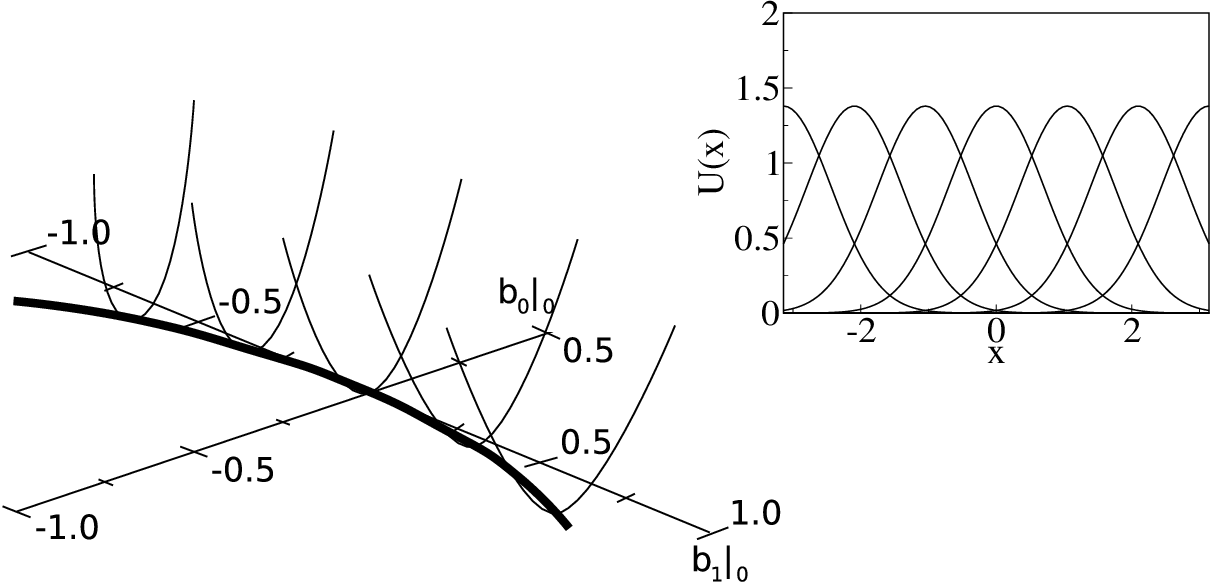,width=7cm}}
\vspace{-0.3cm}
\caption{
The canyon formed by the stationary states 
projected onto the subspace formed by $b_1|_0$ and $b_0|_0$.
Motion along the canyon corresponds 
to the displacement of the bump (inset).}
\end{figure}

The stability of the Gaussian bumps can be studied 
by considering the dynamics of fluctuations. 
Consider the network state $U(x,t)=\tilde U(x|z)+\delta U(x,t)$. 
Then we obtain
\begin{equation}
	\tau\frac{d}{dt}\delta U(x,t)
	=\int dx'F(x,x')\delta U(x',t)-\delta U(x,t),
\label{eq:fluc}
\end{equation}
where the interaction kernel is given by Eq.~(\ref{eq:fxx}) 
in the Appendix.
To compute the eigenfunctions and eigenvalues of the kernel $F(x,x')$, 
we choose the wave functions of the quantum harmonic oscillators 
as the basis, namely, 
\begin{equation}
	v_n(x|z)=\frac{\exp(-\xi^2/2)H_n(\xi)}{\sqrt{(2\pi)^{1/2}an!2^n}},
\label{eq:wavefunction}
\end{equation} 
where $\xi\equiv(x-z)/(\sqrt{2}a)$ 
and $H_n(\xi)$ is the $n^{\rm th}$ order Hermite polynomial function. 
Indeed, the ground state of the quantum harmonic oscillator 
corresponds to the Gaussian bump, 
and the first, second, and third excited states 
correspond to fluctuations in the peak position, width, and skewness 
of the bump respectively (Fig.2). 
As described in the Appendix, 
the eigenvalues of the kernel $F$ are calculated to be
\begin{eqnarray} 
	\lambda_0 & = & 1-(1-k/k_c)^{1/2}, \label{eq:eigen0}\\ 
	\lambda_n & = & 1/2^{n-1}, ~~~\mbox{for}~~ n\ge 1.
\end{eqnarray}
and the first four eigenfunctions of $F$ are
\begin{eqnarray}
	u_0(x|z) & = & v_0(x|z), \\
	u_1(x|z) & = & v_1(x|z), \\
	u_2(x|z) & = & {\frac {\sqrt{2}} 
	{2D_0}}v_0(x|z)+\frac{1-2\sqrt{1-k/k_c}}{D_0}v_2(x|z), \\
	u_3(x|z) & = & \sqrt{\frac{1}{7}}v_1(x,z) + \sqrt{\frac{6}{7}}v_3(x,z),
	\label{eq:eigenf3}
\end{eqnarray}
where $D_0=[(1-2\sqrt{1-k/k_c})^2+1/2]^{1/2}$.
The eigenfunctions of $F$ correspond 
to the various distortion modes of the bump. 
Since $\lambda_1=1$ and all other eigenvalues are less than 1, 
the stationary state is neutrally stable in one component, 
and stable in all other components.
The first two eigenfunctions are particularly important. 
(1) The eigenfunction for the eigenvalue $\lambda_0$ is $u_0(x|z)$, 
and represents a distortion of the amplitude of the bump. 
As we shall see, amplitude changes of the bump 
affect its tracking performance. 
(2) Central to the tracking capability of CANNs, 
the eigenfunction for the eigenvalue 1 is $u_1(x|z)$ 
and is neutrally stable. 
We note that $u_1(x|z)\propto\partial v_0(x|z)/\partial z$, 
corresponding to the shift of the bump position 
among the stationary states. 
This neutral stability is the consequence 
of the translational invariance of the network. 
It implies that when there are external inputs, however small, 
the bump will move continuously. 
This is a unique property associated
with the special structure of a CANN, 
not shared by other attractor models. 
Other eigenfunctions correspond to distortions 
of the shape of the bump. 
For example, the eigenfunction $u_3(x|z)$ 
corresponds to a skewed distortion of the bump.

\begin{figure}
\centerline{\epsfig{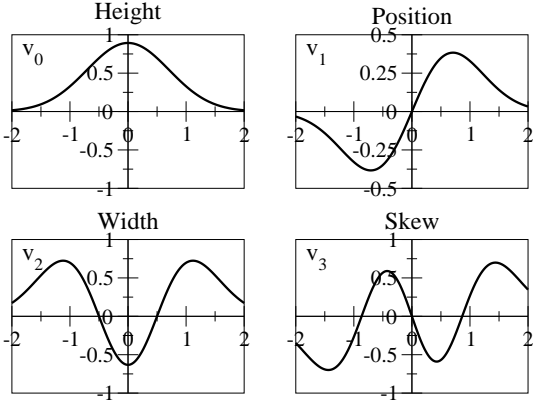}}
\vspace{-0.3cm}
\caption{The first four basis functions of the quantum harmonic oscillators, 
which represent four distortion modes of the network dynamics, 
namely, changes in the height, position, width and skewness 
of a bump state.} 
\end{figure}

It is instructive to consider the energy landscape 
in the state space of a CANN. 
Since $F(x,x')$ is not symmetric, 
a Lyapunov function cannot be derived for Eq.~(\ref{eq:fluc}). 
Nevertheless, for each peak position $z$, 
one can define an effective energy function 
$E|_z=\sum_n(1-\lambda_n)b_n|_z^2/2$, 
where $b_n|_z$ is the overlap between $U(x)-\tilde U(x|z)$ 
and the $n^{\rm th}$ eigenfunction of $F$ centered at $z$. 
Then the dynamics in Eq.~(\ref{eq:fluc}) can be locally described 
by the gradient descent of $E|_z$ 
in the space of $b_n|_z$. 
Since the set of points $b_n|_z=0$ for $n\ne 1$ 
traces out a line with $E|_z=0$ in the state space when $z$ varies, 
one can envisage a canyon surrounding the line 
and facilitating the local gradient descent dynamics, 
as shown in Fig.~1. 
A small force along the tangent of the canyon 
can move the network state easily. 
This illustrates how the landscape of the state space of a CANN 
is shaped by the network structure, 
leading to the neutral stability of the system, 
and how this neutral stability shapes the network dynamics. 

Next, we consider the network dynamics 
in the presence of a weak external stimulus. 
Suppose the neural response at time $t$ is peaked at $z(t)$. 
Since the dynamics is primarily dominated 
by the translational motion of the bump, 
with secondary distortions in shape, 
we may develop a time-dependent perturbation analysis 
using $\{v_n(x|z(t))\}$ as the basis, 
and consider perturbations in increasing orders of $n$. 
This is done by considering solutions of the form
\begin{equation}
	U(x,t)=\tilde U(x|z(t))
	+\sum_{n=0}^\infty a_n(t)v_n(x|z(t)).
\end{equation}
Furthermore, since the Gaussian bump is the steady-state solution 
of the dynamical equation in the absence of external stimuli, 
the neuronal interaction term in Eq.~(\ref{eq:dyn}) 
can be linearized for weak stimuli. 
Making use of the orthonormality and completeness of $\{v_n(x|z(t))\}$, 
we obtain for each order $n$ of perturbation (see Appendix),
\begin{eqnarray}
	\Biggl(\frac{d}{dt}& + &\frac{1-\lambda_n}{\tau}\Biggl)a_n   
	=  \frac{I_n}{\tau}
	-\Biggl[ U_0\sqrt{(2\pi)^{1/2}a}\delta_{n1}  \nonumber \\ 
	& + &\sqrt{n}a_{n-1} -\sqrt{n+1}a_{n+1}\Biggl]
	\frac{1}{2a}\frac{dz}{dt} \nonumber \\ 
	& + &\frac{1}{\tau}\sum_{r=1}^{\infty}\sqrt{\frac{(n+2r)!}{n!}}
	\frac{(-1)^r}{2^{n+3r-1}r!}a_{n+2r},
\label{eq:perturbation}
\end{eqnarray}
where $I_n(t)$ is the projection of the external input 
$I_{\rm ext}(x,t)$ on the $n^{\rm th}$ eigenfunction. 

Determining $z(t)$ by the center of mass of $U(x,t)$, 
we obtain the self-consistent condition (see Appendix)
\begin{equation}
	\frac{dz}{dt}=\frac{2a}{\tau}\!\!\left(\frac
	{I_1+\sum_{n=3,{\rm odd}}^\infty\sqrt{n!!/(n-1)!!}I_n+a_1}
	{U_0\sqrt{(2\pi)^{1/2}a}+\sum_{n=0,{\rm even}}^\infty
	\sqrt{(n-1)!!/n!!}a_n}\right).
\label{eq:peak}
\end{equation}

Eqs.(\ref{eq:perturbation}) and (\ref{eq:peak}) are the master equations 
of the perturbation method.  
We can approximate the network dynamics
up to an arbitary accuracy depending on the choice 
of the order of perturbation.
In practice, low order perturbations already yield very accurate results. 
Below, we consider the network dynamics 
for several kinds of external stimuli.

\begin{figure*}[htb]
\begin{center}
\epsfig{file=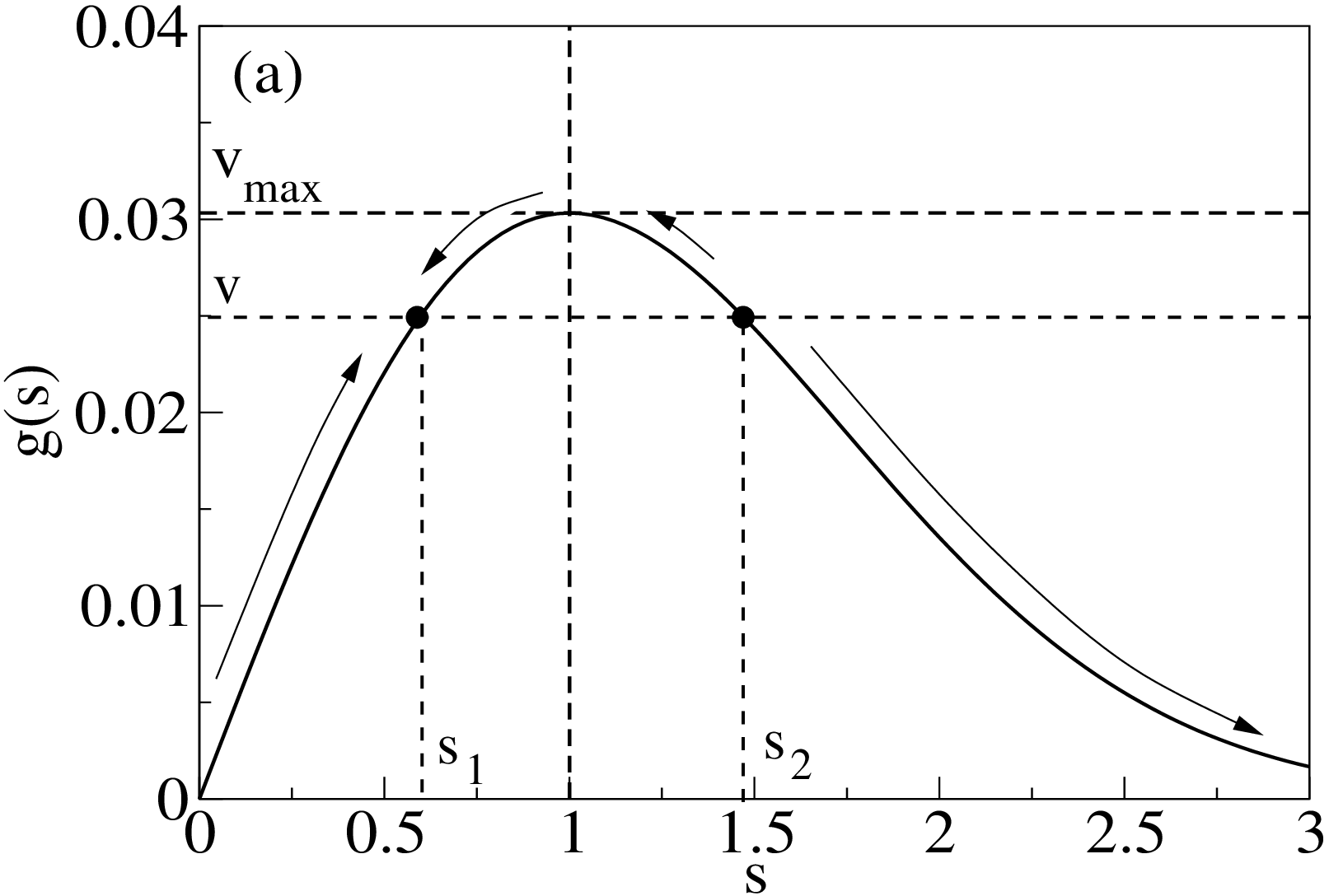,height=3.5cm}
\hspace{0.3cm}\epsfig{file=epl_fig3b.eps,height=3.5cm}
\hspace{0.2cm}\epsfig{file=epl_fig3c.eps,height=3.5cm}
\end{center}
\vspace{-0.3cm}
\caption{(a) 
The function $g(s)$ and the stable and unstable fixed points 
of Eq.~(\ref{eq:lag}). 
(b) The time dependence of the separation $s$ 
starting from different initial values. 
Symbols: simulations with $N=200$. 
Lines: $n=5$ perturbation.
Dashed lines: $s_1$ (bottom) and $s_2$ (top). 
(c) The dependence of the terminal separation $s$ 
on the stimulus speed $v$. 
Symbols: simulations with $N=200$. 
Solid line: prediction of \cite{Wu2}. 
Dashed line: $n=1$ perturbation.
Parameters: $\alpha=0.05$, $a=0.5$, $\tau=1$, $k=0.5$, $\rho=N/(2\pi)$, 
$J=\sqrt{2\pi a^2}$.}
\vspace{-0.2cm}
\end{figure*}

{\it 1) Tracking a stationary stimulus:}
Consider the external stimulus 
consisting of a Gaussian bump, namely, 
$I_{\rm ext}(x,t)=\alpha U_0\exp[-(x-z_0)^2/4a^2]$. 
Perturbation up to the order $n=1$ yields $a_1(t)=0$, 
$[d/dt+(1-\lambda_0)/\tau]a_0=\alpha U_0\sqrt{(2\pi)^{1/2}a}
\exp[-(z_0-z)^2/8a^2]/\tau$, and
\begin{equation}
	\frac{dz}{dt}=\frac{\alpha}{\tau}(z_0-z)
	\exp\left[-\displaystyle\frac{(z_0-z)^2}{8a^2}\right]R(t)^{-1},
\label{eq:ou}
\end{equation}
where $R(t)=1+\alpha\int_{-\infty}^t(dt'/\tau)
\exp[-(1-\lambda_0)(t-t')/\tau-(z_0-z(t'))^2/8a^2]$, 
representing the ratio of the bump height 
relative to that in the absence of the external stimulus ($\alpha=0$). 
Hence, the dynamics is driven by a pull of the bump position 
towards the stimulus position $z_0$. 
When compared with the limiting expression for weak stimuli, 
which was considered in the dynamics of the moving bump in~\cite{Wu2}, 
the pull is reduced by the factor $R(t)$. 
This is due to the increase in amplitude of the bump, 
which slows down its response.

Suppose, in addition to the bumped signal, 
there is also the $I_1$ component consisting of 
a white noise of temperature $T_n$. 
Then the noises shift the bump position randomly. 
In the absence of the bumped signal in the external stimulus ($\alpha=0$), 
the peak position of the bumped response 
undergoes a random walk with $\langle z(t)^2\rangle=2Dt$ 
in the low noise limit, 
where $D\equiv 4aT_n/[(2\pi)^{1/2}U_0^2\tau^2]$ is the diffusion constant. 
In the presence of a weak bumped signal ($\alpha>0$), 
the peak of the bumped response 
experiences a pull which can be written as the gradient of a potential 
$V(z)=-4a^2\alpha\exp[-(z-z_0)^2/8a^2]/\tau$. 
Thus, the probability distribution of $z$ 
can be approximated by a Boltzmann distribution 
$P(z)\propto\exp[-V(z)/D]$ in the low noise limit. 

{\it 2) Tracking a moving stimulus:} 
The tracking performance of a CANN is a key property 
that is believed to have wide applications in neural systems. 
Suppose the stimulus is moving at a constant velocity $v$, 
and the noise is negligible. 
The dynamical equation becomes identical to Eq.~(\ref{eq:ou}), 
with $z_0$ replaced by $vt$. 
Denoting the lag of the bump behind the stimulus by $s=z_0-z$ we have, 
after the transients,
\begin{equation}
	{\frac{ds}{dt}}=v-g(s);\ 
	g(s)\equiv\frac{\alpha se^{-s^2/8a^2}}{\tau}
	\left[1+\frac{\alpha e^{-s^2/8a^2}}{1-\lambda_0}\right]^{-1}.
\label{eq:lag}
\end{equation}
The value of $s$ is determined by two competing factors: 
the first term represents the movement of the stimulus, 
which tends to enlarge the separation, 
and the second term represents the collective effects 
of the neuronal recurrent interactions,
which tends to reduce the lag. 
Tracking is maintained when these two factors match each other, 
i.e., $v=g(s)$; otherwise, $s$ diverges.

Fig.3(a) shows that the function $g(s)$ is concave, 
and has the maximum value of
$g_{\rm max}=2\alpha a/(\tau\sqrt{e})$ at $s=2a$. 
This means that if $v>g_{\rm max}$, 
the network is unable to track the stimulus. 
Thus, $g_{\rm max}$ defines the maximum trackable speed 
of a moving stimulus.
Notably, $g_{max}$ increases with the strength of the external signal 
and the range of neuronal recurrent interactions. 
This is reasonable since it is the neuronal interactions 
that induce the movement of the bump. 
$g_{\rm max}$ decreases with the time constant of the network, 
as this reflects the responsiveness of the network to external inputs.

On the other hand, for $v<g_{\rm max}$, 
there is a stable and unstable fixed point, 
respectively denoted by $s_1$ and $s_2$ in Fig.~3(a). 
When the initial distance is less than $s_2$, 
it will converge to $s_1$. 
Otherwise, the tracking of the stimulus will be lost. 
Figs.~3(b) and (c) show that 
the analytical results of Eq.~(\ref{eq:lag}) 
well agree with the simulation results.

{\it 3) Tracking an abrupt change of the stimulus:} 
Suppose the network has reached a steady state 
with an external stimulus stationary at $t<0$, 
and the stimulus position jumps from 0 to $z_0$ suddenly at $t=0$.
This is a typical scenario 
in experiments studying mental rotation behaviors. 
We first consider the case that the jump size $z_0$ is small 
compared with the range $a$ of neuronal interactions. 
In the limit of weak stimulus, 
the dynamics is described by Eq.~(\ref{eq:ou}) with $R(t)=1$. 
We are interested in estimating the reaction time $T$, 
which is the time taken by the bump 
to move to a small distance $\theta$ from the stimulus position. 
In the limit of low noise, 
the reaction time increases logarithmically 
with the jump size, namely,
$T\approx(\tau/\alpha)\ln(|z_0|/\theta)$.

\begin{figure*}[htb]
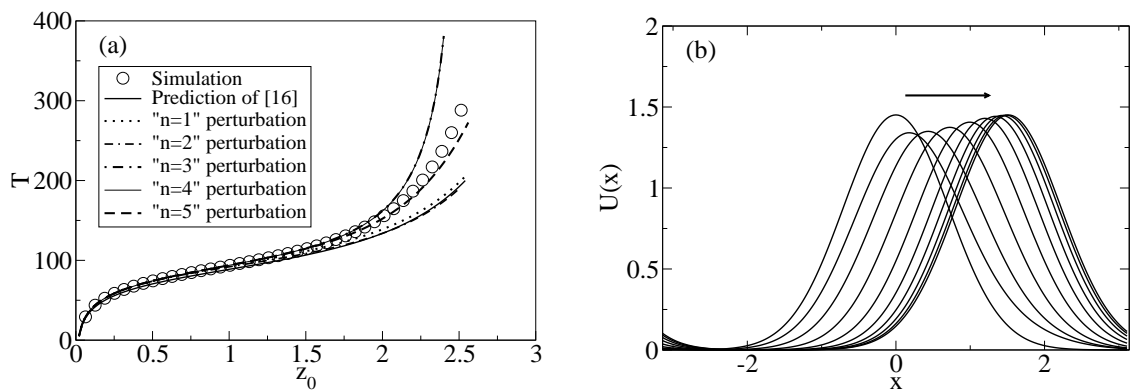

\begin{center}
\vspace{0.5cm}
\epsfig{file=epl_fig4a.eps,width=7cm}
\hspace{0.5cm}
\epsfig{file=epl_fig4b.eps,width=7cm}
\end{center}
\vspace{-0.5cm}
\caption{(a) The dependence of the reaction time $T$ 
on the new stimulus position $z_0$. 
(b) Profiles of the bump between the old and new positions 
at $z_0=\pi/2$ in the simulation. 
Parameters: as in Fig.3.}
\vspace{-0.3cm}
\end{figure*}

When the strength $\alpha$ of the external stimulus is larger, 
improvement using a perturbation analysis up to $n=1$ is required 
when the jump size $z_0$ is large. 
This amounts to taking into account the change of the bump height 
during its movement from the old to new position. 
The result is identical to Eq.~(\ref{eq:ou}), 
with $R(t)$ replaced by
\begin{eqnarray}
	&&R(t)=1+\frac{\alpha}{1-\lambda_0}
	\exp\left[-\frac{(1-\lambda_0)}{\tau}t\right]
	+\alpha\int_0^t\frac{dt'}{\tau}
	\nonumber\\
	&&\times\exp\left[-\frac{(1-\lambda_0)}{\tau}(t-t')
	-\frac{(z_0-z(t'))^2}{8a^2}\right].
\end{eqnarray}
Indeed, $R(t)$ represents the change in height 
during the movement of the bump. 
Contributions from the second and third terms 
show that it is highest at the initial and final positions respectively, 
and lowest at some point in between, 
agreeing with simulation results shown in Fig.~4(b). 
Fig.~4(a) shows that the $n=1$ perturbation 
overcomes the insufficiency of the logarithmic estimate, 
and has an excellent agreement with simulation results 
for $z_0$ up to the order of $2a$. 
We also compute the reaction time up to the $n=5$ perturbation, 
and the agreement with simulations remains excellent 
even when $z_0$ goes beyond $2a$. 
This implies that beyond the range of neuronal interaction, 
tracking is influenced by the distortion 
of the width and the skewed shape of the bump.

To conclude, we have systematically investigated 
how the neutral stability of a CANN 
facilitates the tracking performance of the network, 
a capability which is believed to have wide applications in brain functions. 
Two interesting behaviors are observed, namely,
the maximum trackable speed for a moving stimulus 
and the reaction time for catching up an abrupt change of a stimulus, 
logarithmic for small changes 
and increasing rapidly beyond the neuronal range. 
These two properties are associated with the unique dynamics of a CANN. 
They are testable in practice and can serve as general clues in for checking
the existence of a CANN in neural systems. 
In order to solve the dynamics 
which is otherwise extremely complicated 
for a large recurrent network, 
we have developed a perturbative analysis 
to simplify the dynamics of a CANN. 
Geometrically, it is equivalent to projecting the network state 
on its dominant directions of the state space. 
This method works efficiently and may
be widely used in the study of CANNs. 
Furthermore, the special structure of a CANN may have other applications 
in brain functions, for instance, 
the highly structured state space of a CANN 
may provide a neural basis for encoding
the topological relationship of objects in a feature space,
as suggested by recent psychophysical experiments~\cite{Kourtzi,Wichmann}.
We expect the mathematical framework developed in this study
will be very valuable to explore these issues. 

The tracking dynamics of a CANN has also been studied by other authors. 
In particular, 
Zhang proposed a mechanism of using asymmetrical recurrent interactions 
to generate the bump movement
without the bump shape being distorted \cite{Zhang}. 
Xie \etal  further proposed a double ring network model to implement these
asymmetrical interactions based on the external inputs \cite{Seung}. 
These models work well for the head-direction system. 
Different from their model setting, here we consider the movement of
the bump directly driven by external inputs, 
and hence the shape of the bump is unavoidably distorted.
This setting has also been used by many other authors 
in modeling brain functions (see, e.g., \cite{Dan,Erlhagen}).
We quantify how the distortion of the bump shape 
affects the network tracking performance, and obtain
a new finding on the maximum trackable speed of the network.
We expect these results will enrich
our knowledge on the tracking dynamics of CANNs.

Finally, we would like to remark on the generality of the results 
in this work and their relationships to other studies in the literature.
In order to pursue an analytical solution, we have used 
a divisive normalization to represent the inhibition effect. 
This is different from the Mexican-hat type of
recurrent interactions used by many authors. For the latter, it is often 
difficult to get a closed-form of the network stationary state. 
Amari used a Heaviside function to simplify the neural response, 
and obtained the box-shaped network stationary state \cite{Amari}.
However, since the Heaviside function is not differentiable,
it is difficult to describe the tracking dynamics in the Amari model. 
Truncated sinusoidal functions have been used, 
but it is difficult to use them to describe 
general distortions of the bumps~\cite{Sompolinsky}. 
Here, by using divisive 
normalization and the Gaussian-shaped recurrent interactions, we solve the 
network stationary states and the tracking dynamics analytically.

One may be concerned about the feasibility of the divisive normalization.
Firstly, we argue that neural systems can have resources 
to implement this mechanism \cite{Latham,Grossberg,Heeger}.
Let us consider, for instance, a neural network, in which all excitatory
neurons are connected to a pool of inhibitory neurons. 
Those inhibitory neurons have a time constant much shorter than that of
 excitatory neurons, and they inhibit
the activities of excitatory neurons 
in a uniform shunting way~\cite{Grossberg}, 
thus achieving the effect of divisive normalization. 

Secondly, and more importantly,
the main conclusions of our work are qualitatively indpendent of
the choice of the model. 
This is because our calculation is based on the fact 
that the dynamics of a CANN is 
dominated by the motion mode of position shift of the network state, and 
this property is due to the
translational invariance of the neuronal recurrent interactions, 
rather than the inhibition mechanism.
We have formally proved that for a CANN model, 
once the recurrent interactions are translationally invariant, 
the interaction kernel has a unit eigenvalue
with respect to the position shift mode 
no matter what the inhibition mechanism is 
(we will report this result in a separate publication).

This work is partially supported 
by the Research Grant Council of Hong Kong 
(Grant No. HKUST 603606 and HKUST 603607), BBSRC (BB/E017436/1)
and the Royal Society.

\section{Appendix: Mathematical Derivations}
The interaction kernel involved in Eq.~(\ref{eq:fluc}), 
$F(x,x')$, can be derived as 
\begin{eqnarray}
	&& F(x,x') = {\frac {2} {a\sqrt{\pi}}} 
	e^{-\frac{(x-x')^2}{2a^2}}
	e^{-\frac{(x'-z)^2}{4a^2}}
	\nonumber \\
	&&-{\frac {1+\sqrt{1-k/k_c}} {\sqrt{2\pi}a}}
	e^{-\frac{(x-z)^2}{4a^2}}
	e^{-\frac{(x'-z)^2}{4a^2}}.
\label{eq:fxx}
\end{eqnarray}
In the basis of $\{v_n(x|z)\}$, 
the matrix elements $F_{mn}$ are defined by
\begin{equation}
	F_{mn}=\int\int dxdx' v_m(x|z)F(x,x'|z)v_n(x'|z).
\label{eqA2}
\end{equation}
By doing some mathematics, we have
\begin{eqnarray}
	&&F_{mn} \nonumber \\
	&=& \left \{ \begin{array}{ll} 
	1 - \sqrt{1-k/k_c},  ~~~~~~~\mbox{$m = n = 0$,} \\
	2^{1-n}\sqrt{\frac{n!}{m!}}\frac{(-1)^{\frac{n-m}{2}}}
	{2^{\frac{n-m}{2}}(\frac{n-m}{2})!}, 
	~~~\mbox{$(n-m)/2>0$,} \\
	0,   ~~~~~~~~~~~~~~~~~~~~~~~~~\mbox{otherwise.}\end{array} \right. 
\label{eqA3}
\end{eqnarray}
The result from Eq.~(\ref{eq:eigen0}) to Eq.~(\ref{eq:eigenf3}) follows.

For Eq.~(\ref{eq:perturbation}), since
\begin{equation}
	\tau\frac{d}{dt}v_n(x|z)
	=\left[\sqrt{n+1}v_{n+1}(x|z)-\sqrt{n}v_{n-1}(x|z)\right]
	\frac{1}{2a}\frac{dz}{dt},
\label{eqA4}
\end{equation}
and
\begin{eqnarray}
	 &&\rho\int dx' J(x,x')r(x'|z)-U(x|z) \nonumber \\
	 &\approx& \sum_n a_n \sum_m v_m(x|z)F_{mn}-\sum_n a_n v_n(x|z).
\label{eqA5}
\end{eqnarray}
Substituting into Eq.~(\ref{eq:dyn}) 
and collecting terms, 
we get Eq.~(\ref{eq:perturbation}).

To derive Eq.~(\ref{eq:peak}), 
we take the first moments about the centre of mass,
\begin{equation}
	\sum_{n=1,\rm odd}^\infty\frac{a_n}{\sqrt{n!2^n}}
	\int dx \exp\left(-\frac{x^2}{4a^2}\right)
	xH_n\left(\frac{x}{\sqrt{2}a}\right)=0.
\label{eqA8}
\end{equation}
After integrations, we have
\begin{equation}
	\sum_{n=1,\rm odd}^\infty\sqrt{\frac{n!!}{(n-1)!!}}a_n=0.
\label{eqA9}
\end{equation}
Eq.~(\ref{eq:peak}) follows from combining with Eq.~(\ref{eq:perturbation}).

Further details can be found in \cite{Fung}.


\end{document}